\documentclass[aps,pra,epsfigure,twocolumn,superscriptaddress,nofootinbib]{revtex4-1}
\usepackage[colorlinks=true,linkcolor=blue,urlcolor=blue,citecolor=blue,pdfusetitle]{hyperref}
\usepackage[utf8]{inputenc}
\usepackage{graphicx}
\usepackage{subcaption}
\usepackage[english]{babel}
\usepackage{amsmath}
\usepackage{graphicx,epstopdf}
\usepackage{float}
\usepackage{blindtext}
\usepackage{comment}
\usepackage[table,xcdraw]{xcolor}
\usepackage{lipsum}
\usepackage{amsfonts}
\usepackage{bbm}
\usepackage{amssymb}
\usepackage{enumerate}
\usepackage{color}
\usepackage{latexsym}
\usepackage{times,txfonts}

\newcommand{\Ccal}{\mathcal{C}}
\newcommand{\Dcal}{\mathcal{D}}

\newcommand{\Hcal}{\mathcal{H}}

\newcommand{\Wcal}{\mathcal{W}}

\newcommand{\1}{\mathbbm{1}}

\newcommand{\dbar} {\ensuremath{\,\mathchar'26\mkern-12mu d}}

\begin{document}

\title{Quantum caloric effects}

\author{C. Cruz}
\email{clebson.cruz@ufob.edu.br}
\affiliation{Quantum Information Group, Center for Exact Sciences and Technologies, Federal University of Western Bahia - Campus Reitor Edgard Santos, Rua Bertioga, 892, Morada Nobre I, 47810-059 Barreiras, Bahia, Brazil.}

\author{J. S. Amaral}
\email{jamaral@ua.pt}
\affiliation{Department of Physics and CICECO, University of Aveiro, Universitary Campus of Santiago, 3810-193 Aveiro, Portugal}

\author{Mario Reis}
\email{marioreis@id.uff.br}
\thanks{Permanent address: Federal Fluminense University, Brazil; Temporary address: University of Aveiro, Portugal.}
\affiliation{Department of Physics and CICECO, University of Aveiro, Universitary Campus of Santiago, 3810-193 Aveiro, Portugal}
\affiliation{Institute of Physics, Federal Fluminense University, Av. Gal. Milton Tavares de Souza s/n, 24210-346 Niterói, Rio de Janeiro, Brazil.}

\date{\today}

\begin{abstract}
Quantum thermodynamics aims to explore quantum features to enhance energy conversion beyond classical limits. While significant progress has been made, the understanding of caloric potentials in quantum systems remains incomplete. In this context, this study focuses on deriving general expressions for these caloric potentials, by developing a quantum Maxwell relationship obtained from a thermal average form of the Ehrenfest theorem. Our results recover the classical cases and also reveal that the isothermal entropy change can be related to genuine quantum correlations in the system. Thus, this work {aims to contribute to the understanding of the caloric behavior of quantum systems and their potential implication in caloric devices.}

\end{abstract}

\maketitle

\section{Introduction}
 
Thermodynamics is one of the fundamental theories in physics because of its broad applicability, providing accurate predictions for classical systems. Even in the disruptive realm of quantum mechanics, it is impossible for heat to spontaneously flow from a cold reservoir to a hot one. Consequently, the efficiency of a heat engine cannot exceed the Carnot limit. Equilibrium thermodynamics does not impose restrictions on the rate at which energy can be converted into heat or work, sparking interest in using quantum systems as thermodynamic devices \cite{myers2022quantum,deffner2019quantum,Horodecki2011Fundamental,Seifert2012Stochastic,campaioli2023colloquium}. The potential for quantum systems to surpass classical limitations in energy conversion rates has led to ongoing research in the emerging field of Quantum Thermodynamics \cite{Gemmer:Book,deffner2019quantum,binder2018thermodynamics,vinjanampathy2016quantum,myers2022quantum,Vinjanampathy2015Quantum,Millen2015Perspective}. By harnessing unique properties of quantum mechanics, such as superposition and entanglement, scientists aim to develop more efficient and advanced thermodynamic devices for various applications \cite{myers2022quantum,campaioli2023colloquium,Campaioli2018,campaioli2023colloquium,PhysRevLett.130.150201}.

Despite significant progress, the understanding of caloric potentials in quantum thermodynamic systems remained unexplored. Traditional thermodynamic frameworks for studying the caloric effects in advanced quantum materials \cite{reis2020magnetocaloric,Franco2018Magnetocaloric,Smith2012Materials} do not fully capture the nuances of its quantum features.% introduced by quantum correlations. 
 This gap in the literature highlights the need for a deeper exploration of quantum caloric effects. Addressing this need could lead to groundbreaking advancements in how we manipulate energy conversion processes at the quantum level, shedding light on fields like the energetic costs of quantum processes \cite{PhysRevLett.124.130601,Linpeng2022Energetic}, quantum computing \cite{Jaschke2022Is,Arrachea2022Energy}, and nanoscale engines \cite{PhysRevLett.112.030602,PhysRevLett.126.180605}.

In this scenario, this article introduces the quantum caloric effect, examining the caloric potentials of quantum systems. By deriving a Maxwell relationship adapted for quantum thermodynamics, obtained from a thermal average form of the Ehrenfest theorem, we develop general expressions for: (i) isothermal entropy change, which depends on the thermal average of the Hamiltonian's derivative with respect to the external excitation, and (ii) adiabatic temperature change, which also depends on the variance of the energy spectra. {The derivation of these key thermodynamic quantities is built on the primary assumption that the quantum systems of interest are in thermal equilibrium, and we ensure a commutative relationship between the Hamiltonian and the density operator. The equilibrium condition simplifies the mathematical development, ensuring that quantum definitions of heat and work recover their classical counterparts bridging both quantum and classical frameworks.}

In this regard, we apply the proposed framework to the magnetic scenario, demonstrating that it not only can recover the classical cases but also expands our understanding of the caloric effects, revealing that the isothermal entropy change depends on genuine quantum features of the system, such as quantum discord. Therefore, {the proposed approach offers a perspective on how quantum systems can be manipulated to influence cooling and heating processes, complementing classical thermodynamic approaches and providing insights into the relationship between classic caloric effects and quantum thermodynamics. The main contribution of this paper is demonstrating that, considering the thermal equilibrium in which quantum systems obey the same thermodynamic constraints as their classical analogs, its caloric behavior can be significantly altered by quantum effects beyond the consideration of discrete energy levels. These include quantum correlations, which influences heat exchange mechanisms in ways not captured by the classical framework.}

{The paper is structured following the theoretical development, which includes the derivation of a Maxwell relationship adapted to quantum thermodynamics, grounded in the thermal analog of the Ehrenfest theorem in Section \ref{sec2}. Section \ref{sec3} discusses quantum isothermal and adiabatic processes, focusing on quantum caloric potential related to the isothermal variation of entropy and the adiabatic temperature changes. Section \ref{sec4} applies the developed framework to magnetic scenarios, connecting the results to the classical case and surpassing it by exploring the role of quantum correlations on heat exchange in a dinuclear metal complex. The article is concluded in Section \ref{sec5}, summarizing the contribution to the literature, emphasizing the advancements in studying caloric effects through quantum thermodynamics and the potential technological applications of the proposed approach.}

\section{Theoretical Foundations}\label{sec2}

{It is worth highlighting that a significant portion of this section reviews and extends well-established concepts in thermodynamics and statistical mechanics, adapted to the quantum regime.}
Let us start considering a general Hamiltonian in the energy basis $\{|n\rangle\}$, expressed as:
\begin{align}\label{eq:Hamiltonian}
\Hcal(\lambda) = \sum_n E_n(\lambda) |n\rangle \langle n|,
\end{align}
where $\lambda$ represents a parameter that is able to change the energy levels of the system, such as magnetic fields \cite{bouton2021quantum} or magnetic couplings \cite{cruz2022}, for instance. %or pressure
{The study makes an assumption that the system is in a state of thermal equilibrium\footnote{{Without this assumption, the system would exhibit non-equilibrium dynamics, where the population of energy levels would evolve in time, introducing additional terms and complexities requiring a time-dependent quantum thermodynamics, which are beyond the scope of this work.}},} thus the corresponding density operator of the system, can be written in the same energy basis and is given by $\rho(\lambda,T) = \sum_n p_n(\lambda, T) |n\rangle \langle n|$, where $p_n(\lambda, T) = e^{-E_n(\lambda) / k_B T}/Z(\lambda, T)$ represents the Boltzmann weights of each state and {allowing for a well-defined partition function $Z(\lambda, T)$.} 

The internal energy is computed through the thermal average of the Hamiltonian:
\begin{align}
U(\lambda, T) = \langle \mathcal{H}(\lambda)\rangle = \operatorname{Tr}[\rho(\lambda, T) \mathcal{H}(\lambda)] = \sum_n p_n(\lambda, T) E_n(\lambda),
\end{align}
leading to the quantum analog of the first law of thermodynamics \cite{quan2009quantum,Quan:07}:
\begin{align}\label{eq:du}
dU = \dbar \Wcal + \dbar Q = \sum_n \left[p_n(\lambda, T) dE_n(\lambda) + E_n(\lambda) dp_n(\lambda, T)\right],%\sum_n \left[\underbrace{p_n(\lambda, T) dE_n(\lambda)}_{\delta W} + \underbrace{E_n(\lambda) dp_n(\lambda, T)}_{\delta Q}\right],
\end{align}
where $\dbar \Wcal = \sum_n p_n(\lambda, T) dE_n(\lambda)$ and $\dbar Q = \sum_n E_n(\lambda) dp_n(\lambda, T)$ denote the quantum work and quantum heat, respectively, as defined by Alicki \cite{Alicki:18}. {In this regard, $\lambda$ can be interpreted by the working parameter of the system, since the process of performing work on a system can be described in a general way by changing this parameter in the Hamiltonian \cite{deffner2019quantum}.}

{The differential form of the first law of thermodynamics given by equation \eqref{eq:du} can be used as a foundation, providing continuity with classical approach of caloric effects \cite{DEOLIVEIRA201089}, while allowing us to introduce quantum aspects systematically. In the following, we introduce the framework that unifies the theoretical aspects of caloric effects and the quantum thermodynamics through the development of a thermal analog of the Ehrenfest theorem. This result not only ensures the consistency of quantum definitions of work and heat given by Alicki \cite{Alicki:18} with their classical counterparts, but also facilitates the derivation of Maxwell-type relations for the Helmholtz free energy. Consequently, this approach can establish a formal bridge between microscopic quantum features and macroscopic caloric phenomena, offering innovative insights into the caloric properties and energy transfer processes in quantum systems.}

\subsection{Thermal Analogs of the Ehrenfest Theorem}
{For the system in thermal equilibrium, the Hamiltonian $H(\lambda)$ and the density operator $\rho(\lambda, T)$ commute, i.e., $[H(\lambda), \rho(\lambda, T)] = 0$. This simplifies the mathematical formulation, as it ensures that quantum definitions of heat and work recover their classical counterparts \cite{deffner2019quantum,PhysRevE.102.062152}.} 
Nonetheless, the generalized classical form of the first law of thermodynamics can be expressed as follows \cite{quan2009quantum}:
\begin{equation}
dU = \sum_i Y_i dq_i + TdS,
\end{equation}
where 
\begin{equation}
Y_i = -\frac{\dbar \Wcal}{dq_i}
\label{eq:genforce}
\end{equation}
represents the generalized force and $q_i$ the corresponding generalized coordinate \cite{quan2009quantum}. The index $i$ in equation \ref{eq:genforce} denotes the different types of work performed on or by the system, such as mechanical, electric, or magnetic work, for instance \cite{Franco2018Magnetocaloric}. Considering that the system eigenvectors do not depend on the work parameter (equation \ref{eq:Hamiltonian}), and using equations \ref{eq:du} and  \ref{eq:genforce} with $\lambda = q_i$, we find:
\begin{align}\label{eq:ehrenfest0}
Y(\lambda, T) &= -\frac{\dbar \Wcal}{d \lambda} = -\sum_n p_n(\lambda, T) \frac{dE_n(\lambda)}{d\lambda}~,
\end{align}
where, according to the quantum definition of work used, only energy displacement occurs. Thus, for any quantum system in thermal equilibrium, where $\left[ \Hcal(\lambda), \rho (\lambda) \right] = 0$,  the generalized force represents the equilibrium thermal average of the first derivative of the Hamiltonian with respect to the working parameter $\lambda$ {(for details, see Appendix \ref{ehrenfest})}. 
\begin{align}\label{eq:ehrenfest}
Y(\lambda, T) =- \left\langle \frac{d \Hcal (\lambda)}{d \lambda}\right\rangle,
\end{align}
In other words, the above equation can be interpreted as the \textit{thermal analogous of the Ehrenfest theorem} \cite{Sakurai:Book}. {Thus, from equation \ref{eq:ehrenfest}, one can establish that the generalized force associated with a working parameter $\lambda$ in a quantum system can be interpreted as the thermal average of the derivative of the Hamiltonian with respect to that parameter.}

{In this regard, the thermal analog of the Ehrenfest theorem confirms that the concept of generalized force and work in quantum systems can be directly related to the thermal averages in the system. This provides a conceptual and formal bridge between quantum and classical regimes, allowing classical insights to be extended to describe quantum systems. Applying this approach within an equilibrium thermodynamics framework provides a detailed description of the relationships between a given external work parameter and the thermal properties of the quantum system under consideration. This framework can serve as a guide for manipulating quantum systems to optimize energy transfer and enhance efficiency. In the following, we illustrate how this theorem can be applied to derive a Maxwell relationship within the framework of quantum thermodynamics and demonstrate its use in evaluating the caloric potential of quantum systems in thermal equilibrium.}

\subsection{Maxwell relationship in the framework of quantum thermodynamics}
It is also important to highlight that, regarding the quantum definition of heat, for the system at thermal equilibrium, $\dbar Q = \sum_n E_n(\lambda) dp_n(\lambda, T) = TdS$, where $S$ is the Shannon entropy of the system \cite{PhysRevE.102.062152}: $S(\lambda,T)= -k_B\sum_n p_n(\lambda,T) \ln{\left[ p_n (\lambda,T)\right]}$.
    
On the other hand, the free energy and its differential are described in \cite{reis2013fundamentals} as follows:
\begin{align}
F = U - TS \quad \Rightarrow \quad dF = \dbar \Wcal - SdT.
\end{align}
Considering Alicki's definition of quantum work \cite{Alicki:18} and that \( S = S(\lambda, T) \), which is due to the Shannon entropy, it is possible to derive:
\begin{align}
dF(\lambda, T) = \sum_n p_n(\lambda, T) dE(\lambda) - S(\lambda, T) dT
\end{align}
and, consequently:
\begin{align}\label{dF1}
dF(\lambda, T) = \left\langle \frac{d \Hcal (\lambda)}{d \lambda}\right\rangle d\lambda - S(\lambda, T) dT.
\end{align}
Given the dependence \( F = F(\lambda, T) \), we can also express:
\begin{align}\label{dF2}
dF(\lambda, T) = \left(\frac{\partial F}{\partial \lambda}\right)_T d\lambda + \left(\frac{\partial F}{\partial T}\right)_\lambda dT.
\end{align}
By comparing equations \ref{dF1} and \ref{dF2}, we find:
\begin{align}
S = -\left(\frac{\partial F}{\partial T}\right)_\lambda \quad \text{and} \quad \left\langle \frac{d \Hcal (\lambda)}{d \lambda}\right\rangle = \left(\frac{\partial F}{\partial \lambda}\right)_T.
\end{align}
From the above equations, we can obtain the quantum analogs of Maxwell relations for this free energy, given by:
\begin{equation}\label{eq:mx}
\left(\frac{\partial S}{\partial \lambda}\right)_T= - \frac{\partial }{\partial T}\left\langle \frac{d \Hcal (\lambda)}{d \lambda}\right\rangle_\lambda
\end{equation}

Thus, the thermal Ehrenfest theorem, equation \ref{eq:ehrenfest}, leads to the quantum analogs of Maxwell's relations for the free energy, equation \ref{eq:mx} - which, on its turn, represents a significant step toward understanding the thermal behavior of the system at the quantum level. Therefore, this result offers insight into the relationship between thermal properties and quantum behavior. It sheds light on the study of caloric effects in quantum systems and opens up avenues for further exploration of quantum thermodynamics by assessing the potential of quantum systems as caloric devices.

\section{Quantum caloric effects}\label{sec3}

\subsection{Quantum isothermal entropy change}

\begin{figure}
   \centering
\includegraphics[width=\linewidth]{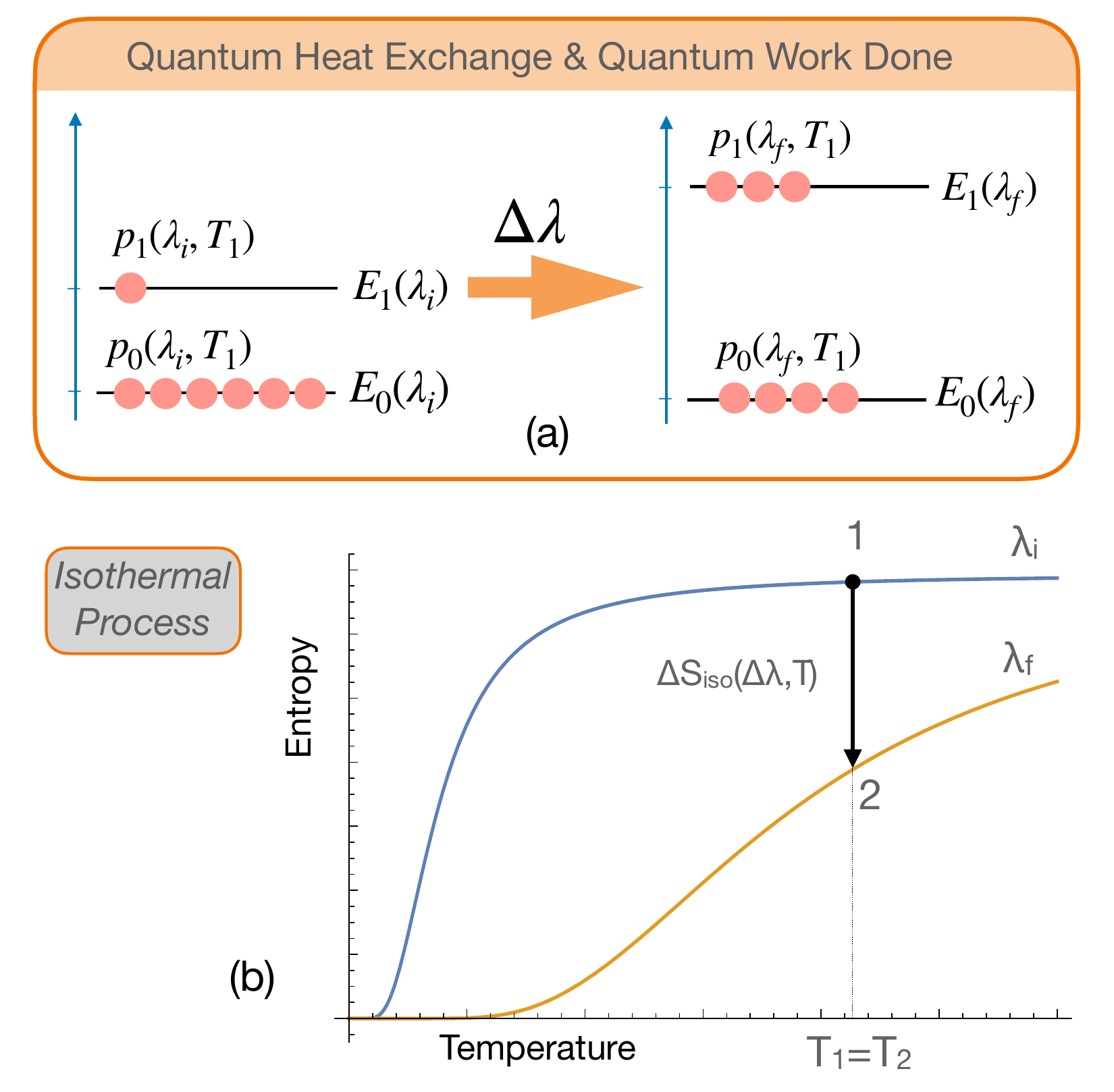}
    \caption{(a) Energy levels \( E_n(\lambda) \) and occupation probabilities \( p_n(\lambda,T) \) of a quantum system during an isothermal process with stable temperature and external stimulus \( \Delta\lambda \). (b) Entropy vs. temperature showing the system's change from \( \lambda_i \) (point 1) to \( \lambda_f \) (point 2) at constant temperature \( T_1 = T_2 \). The heat exchange is quantified by \( T_1 \Delta S_{iso}(\Delta\lambda,T) \).}
    \label{fig:DS}
\end{figure}
{The fundamental principles underlying the derivation of caloric potentials are grounded on the traditional thermodynamic framework, where Maxwell relations play a central role in connecting different thermodynamic quantities \cite{Franco2018Magnetocaloric,reis2020caloric}.}
For this quantum process, analogous to its classical counterpart, the system keeps the temperature stable, while there are simultaneous changes in the energy levels $E_n(\lambda)$ and occupation probabilities $p_n(\lambda,T)$ of the system \cite{Quan:07}. This effect can be seen in Figure \ref{fig:DS}(a), which displays the energy levels of a hypothetical system and the corresponding occupancy. As the process unfolds, both heat and work take place, driven by an external stimulus $\Delta\lambda$. The entropy versus temperature is depicted in Figure \ref{fig:DS}(b) and illustrates the system going from points (1) and (2), which are located on the entropy lines for parameters $\lambda_i$ and $\lambda_f$, respectively, while maintaining a steady temperature $T_1 = T_2$. During the change from step $1\rightarrow 2$, the working substance initiates at point 1 with an energy spectrum $E_n(\lambda_i)$ and reaches a final energy spectrum $E_n(\lambda_f)$. Throughout the isothermal process, heat is exchanged as a result of shifts in the occupation probabilities - and this amount of heat exchanged $T_1\Delta S_{iso}(\Delta\lambda,T)$ is characterized by the quantum isothermal entropy change $\Delta S_{iso}(\Delta\lambda,T)$. Considering the Maxwell relationship for quantum thermodynamics, equation \ref{eq:mx}, we can obtain the total amount of isothermal entropy change:
\begin{equation}
\Delta S_{iso}(\Delta \lambda, T)=-\int_{\lambda_i}^{\lambda_f}\frac{\partial }{\partial T}\left\langle \frac{d \Hcal (\lambda)}{d \lambda}\right\rangle_\lambda d \lambda,
\label{DeltaS_iso}
\end{equation}

\subsection{Quantum adiabatic temperature change}

\begin{figure}
  \centering
\includegraphics[width=\linewidth]{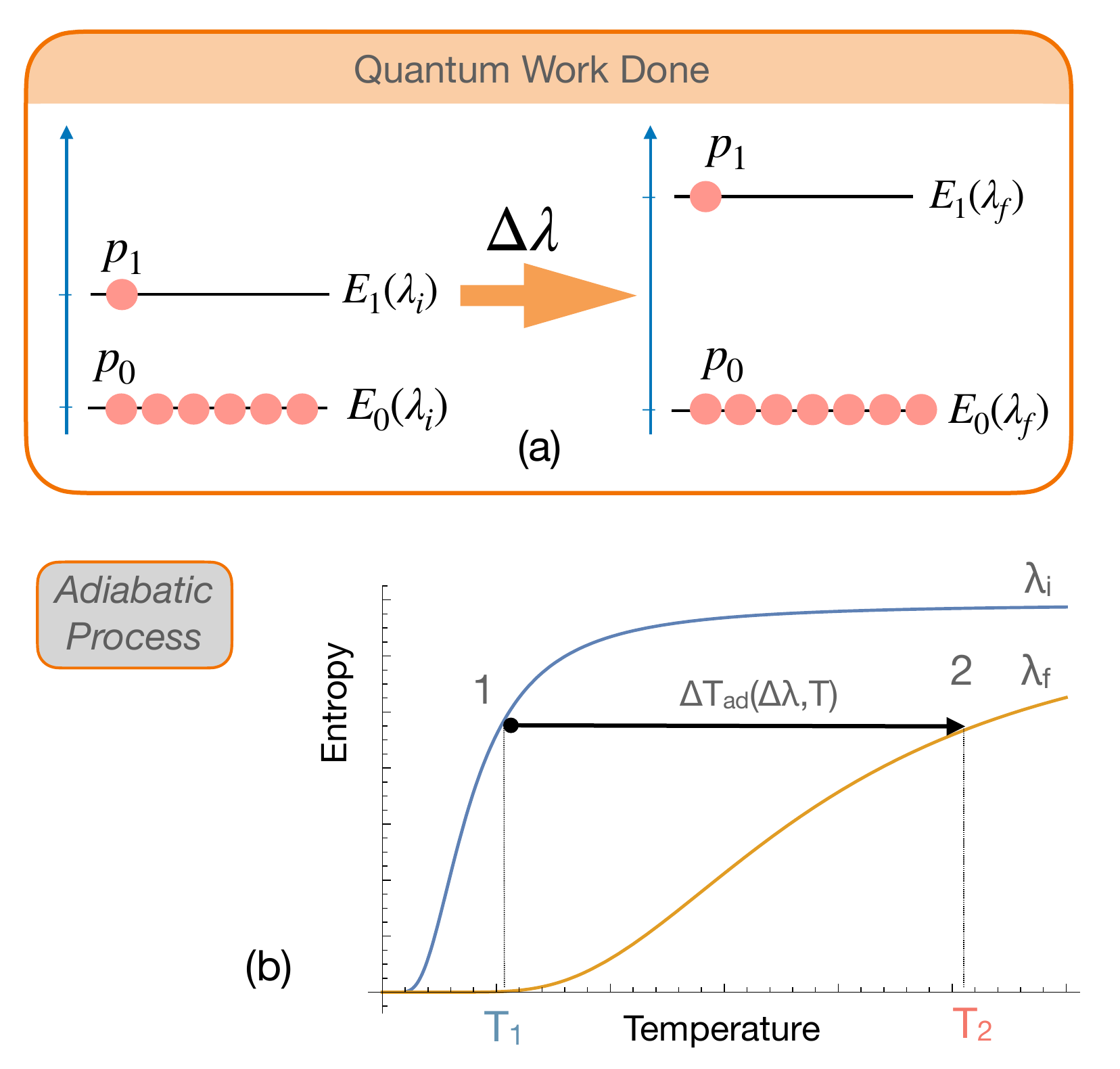}
    \caption{(a) Energy levels \( E_n(\lambda) \) and hypothetical occupancies \( p_n \) for an adiabatic process with no heat transfer (\(dQ = 0\)). External excitation changes the energy levels while occupation probabilities remain constant. (b) Entropy vs. temperature presenting the adiabatic change from \(\lambda_i\) (point 1) at temperature \( T_1 \) to \(\lambda_f\) (point 2) at temperature \( T_2 \). This process highlights the temperature change \( T_1 \to T_2 \) with no change in occupation probabilities.}
    \label{fig:DT}
\end{figure}

One challenge on quantum thermodynamics is determining the inverse function of entropy to express adiabatic temperature changes \cite{Skelt2020,PhysRevA.104.L030202}. To overcome this, we derived
a general expression for the adiabatic temperature change in a quantum process at thermal equilibrium using thermodynamic relations and quantities, as has been done in classical caloric effects \cite{reis2020magnetocaloric,Franco2018Magnetocaloric,Smith2012Materials}.

This process is characterized by the absence of heat transfer between the working substance and the thermal reservoir, resulting in \(\dbar Q=0\). According to the first law of quantum thermodynamics, equation \ref{eq:du}, we can express this as \(dU = \dbar W = \sum_n p_n dE_n\). This phenomenon is illustrated in Figure \ref{fig:DT}(a), which depicts the energy levels and the corresponding hypothetical occupancies of the system. Since there is no heat transfer and only work is performed, the energy levels are changed through external excitation, while the probabilities of occupation for these levels remain constant. Turning our focus to the entropy versus temperature diagram shown in Figure \ref{fig:DT}(b), the process goes from point (1) to point (2). These points are on the entropy curves corresponding to parameters \(\lambda_i\) and \(\lambda_f\), and the adiabatic change involves a temperature change from \(T_1\) to \(T_2\). Specifically, the working substance begins at temperature \(T_1\) with an energy spectrum \(E_n(\lambda_i)\). As work is performed, the system achieves thermal equilibrium at temperature \(T_2\) with a new energy spectrum \(E_n(\lambda_f)\). Throughout this process, no heat is exchanged, as the occupation probabilities remain constant.

The thermodynamics of this process can be described considering \( S = S(\lambda, T) \), as before. Its differential reads as:
\begin{equation}
d S=\left(\frac{\partial S}{\partial T}\right)_\lambda d T+\left(\frac{\partial S}{\partial \lambda}\right)_T d \lambda,
\end{equation}
and the adiabatic condition \( dS=0 \) imposes:
\begin{equation}\label{eq:ger}
\left(\frac{\partial S}{\partial T}\right)_\lambda d T=-\left(\frac{\partial S}{\partial \lambda}\right)_T d \lambda.
\end{equation}
To determine $dT$ we need to further explore the left side of the above equation, considering the right side was already determined in equation \ref{eq:mx}. Thus, for a quantum isochoric process the energy spectra $E_n(\lambda)$ is constant and, therefore, $\dbar W=0$ and $\dbar Q= dU$. Based on this, the specific heat at constant $\lambda$ reads as:
\begin{align}
  \nonumber C_\lambda (T) & =\frac{dQ}{dT}=T\frac{\partial S}{\partial T}=\frac{\partial U(\lambda, T)}{\partial T}\\
  &=\frac{1}{k_B T^2}\left[\langle \Hcal (\lambda)\rangle^2-\langle \Hcal (\lambda)^2\rangle\right]
  =-\frac{var[\Hcal (\lambda)]}{k_B T^2}
\end{align}
where $var[x]=\langle x^2\rangle-\langle x\rangle^2$ represents the variance. Thus, we can write:
\begin{equation}\label{eq:dsdt}
-\left(\frac{\partial S}{\partial T}\right)_\lambda=\frac{var[\Hcal (\lambda)]}{k_B T^3}
\end{equation}
By substituting equations \ref{eq:mx} and \ref{eq:dsdt} into \ref{eq:ger} lead to the quantum adiabatic temperature change:
\begin{equation}\label{eq:DeltaT_ad}
\Delta T_{ad}(\Delta\lambda, T)=-k_B T^3 \int_{\lambda_i}^{\lambda_f} \frac{d\lambda}{var[\Hcal (\lambda)]}\frac{\partial }{\partial T}\left\langle \frac{d \Hcal (\lambda)}{d \lambda}\right\rangle_\lambda.
\end{equation}

{
In this regard, in the high-temperature limit, the system will experience the convergence between quantum and classical description since the variance of the energy levels becomes negligible relative to thermal energy. Thus, from equation \eqref{eq:DeltaT_ad}, the behavior of the adiabatic temperature change in the high-temperature regime follows trends observed in classical caloric effects, in which the internal interactions given by the Hamiltonian model weaken compared to the thermal fluctuation, leading to a reduction in the caloric effect and, consequently, a decrease in $\Delta T_{\text{ad}}$.}

{In summary, the proposed approach for the caloric potentials not only extends the existing concepts but primarily serves to generalize classical insights into the heat exchanges into quantum thermodynamics. The derivation of the quantum caloric potentials, as expressed in Equations \eqref{DeltaS_iso} and \eqref{eq:DeltaT_ad}, follows the standard thermodynamic framework analogs to its classical counterpart \cite{DEOLIVEIRA201089}. However, its key distinction lies in the role of the microscopic Hamiltonian model, which fundamentally differentiate quantum and classical caloric approaches. Nonetheless, it sets the stage for exploring implications of quantum caloric effects never introduced before in the quantum thermodynamics. By structuring our discussion around these foundations, we show in the following section that our framework remains consistent with both classical thermodynamics and modern quantum thermodynamics perspectives, thus reinforcing its validity and applicability.
}

\section{Applications} \label{sec4}

The quantum caloric effects previously discussed have broad and significant implications, offering a comprehensive framework for evaluating the caloric potential of any quantum system in equilibrium with a reservoir. These effects are influenced by changes in the energy spectrum and/or occupancy probability due to variations in a Hamiltonian parameter, denoted by $\lambda$. To illustrate this concept, in this section, we will explore two examples of applications where the Hamiltonian parameter is (i) a magnetic field and (ii) the Heisenberg exchange integral. 

\subsection{Connections with classical results}

The model developed in this work is general. Therefore, we consider a Hamiltonian of the form $\Hcal = \Hcal_{int} + \Hcal_{Zee}$%$\Hcal = \Hcal_{\text{int}} - BM$, 
where $\Hcal_{\text{int}}$ represents any interaction within the system, and $\Hcal_{Zee}= -BM$ denotes the Zeeman energy. By setting $\lambda = B$, we find:
\begin{align}\label{eq:mag1}
\left\langle \frac{d \Hcal}{d B} \right\rangle = -M(T, B).
\end{align}
Substituting this result into equation \ref{DeltaS_iso}, we recover the classical magnetic entropy change \cite{reis2020magnetocaloric,Franco2018Magnetocaloric,Smith2012Materials}.

The adiabatic temperature change, however, has distinct features that require careful analysis of both the classical and quantum versions. Considering equation \ref{eq:DeltaT_ad}, the quantum case is given by:
\begin{equation}\label{eq:DeltaT_ad2}
\Delta T_{\text{ad}}^q(\Delta B, T) = \int_{0}^{B} \frac{T}{c_{B}(T)} \frac{\partial M(T, B)}{\partial T} \, dB,
\end{equation}
where $c_{B}(T) = -\mathrm{var}[\Hcal]/k_B T^2$ represents the magnetic contribution to the specific heat at constant $B$. Thus, this quantum adiabatic temperature change can also be interpreted as depending on energy fluctuations due to thermal effects in the presence of a constant magnetic field. However, there is an additional constraint: during the excitation, the system changes its energy spectra while maintaining constant the corresponding populations. This condition can be expressed as $p_n(B_i,T_i) = p_n(B_f,T_f)$ \cite{Quan:07}. Considering $\beta E_n(B) \ll 1$, we can simplify this to $Z(B_i,T_i) \approx Z(B_f,T_f)$. With this result in hand, and considering the general Hamiltonian $\Hcal = \Hcal_{\text{int}} - BM$, as well as the high-temperature expansion for the magnetization $M = BC/T$ (where $C$ is the Curie constant), we obtain the reversibility condition that must be satisfied to verify the equation \ref{eq:DeltaT_ad2}:
\begin{equation}\label{eq:adcon}
    \frac{\mu_BB_i}{k_BT_i}=\frac{\mu_BB_f}{k_BT_f}.
\end{equation}
This is a sufficient and necessary condition to ensure that the total entropy remains constant during the process \cite{Quan:07}. 
On the other hand, the classical case, which is well known from the literature \cite{reis2020magnetocaloric,Franco2018Magnetocaloric,Smith2012Materials}, is given by:
\begin{equation}\label{eq:DeltaT_ad3}
\Delta T_{\text{ad}}^c(\Delta B, T) = \int_{0}^{B} \frac{T}{c_{\text{tot}}(T)} \frac{\partial M(T, B)}{\partial T} \, dB,
\end{equation}
where $c_{\text{tot}}(T) = c_l(T) + c_s(B, T)$ is the total specific heat, consisting of a lattice contribution $c_l(T)$, which depends only on the temperature, and a magnetic contribution $c_m(B,T)$, which depends on both the magnetic field and the temperature.

Note that these two adiabatic temperature changes have fundamental differences. Both are defined for zero total entropy change. For the classical case, increasing the magnetic field results in a decrease in the magnetic entropy. To maintain constant total entropy, another entropy contribution must increase. This is typically the lattice entropy, which rises to compensate for the decrease in the magnetic entropy. By increasing the lattice entropy, the system's temperature rises. The mechanism is different for the quantum adiabatic entropy change. By changing the magnetic field, only the energy levels change, while the populations remain constant. For this case, only work is done during this process, with no heat being produced - as a consequence, there is no need of extra term to the entropy, as in the classical term, and magnetic entropy (equal to the total entropy) is constant. 

In the scenario where $\lambda = B$, we observe a return to the classical case due to the conjugate pair magnetization - magnetic field. When the magnetic field is treated as the Hamiltonian parameter, its conjugate variable, magnetization, becomes particularly significant. 
This recovery of the classical scenario can be attributed to the Zeeman effect, which describes the interaction of individual spins with an external magnetic field and is intrinsically linked to the local properties of these spins, underscoring the role of magnetization in characterizing the local properties of individual spins within a solid \cite{wiesniak2005magnetic}.

\subsection{Connections with quantum correlations}

As discussed above, the pair magnetization - magnetic field relates to local properties of the system due to the interaction of individual spins with the external magnetic field. This interaction allows the quantum caloric effect to revert to the classical case. Conversely, properties like magnetic susceptibility are tied to spin-spin correlation functions, revealing quantum correlations within the system \cite{cruz2023quantum,cruz2020quantifying,cruz2022quantum,balazadeh2020quantum,ghannadan2023magnetic,strevcka2024room}. In this regard, since the results presented in this paper are applicable to any quantum system in thermal equilibrium, we can evaluate the quantum caloric effect based on different working parameters in the magnetic scenario and assess the impact of system's quantum features on its caloric potential. Thus, let us consider the internal interaction as described by the isotropic Heisenberg interaction: 
\begin{equation}
    \Hcal_{int} = J \sum_{i \neq j}\vec{S}_{i}\cdot\vec{S}_{j}~.
\end{equation}
By setting the work parameter as $\lambda = J$, we obtain:
\begin{align}
\left\langle\frac{d \Hcal}{d J}\right\rangle&=\sum_{i \neq j}\left\langle\vec{S}_i \cdot \vec{S}_j\right\rangle = \sum_{\alpha}\sum_{i \neq j}  \Ccal_{i,j}^{(\alpha)}(J, T)~,\label{eq:ehrenfest_correlation}
\end{align}
where $ \Ccal_{i,j}^{(\alpha)}(J, T) = \left\langle{S}_i^{(\alpha)}{S}_j^{(\alpha)}\right\rangle$ is the pairwise spin-spin correlation function \cite{yurishchev2011quantum,aldoshin2014quantum,cruz2016carboxylate,cruz2020quantifying}, with $\alpha = \{ x,y,z\}$. %\textcolor{red}{VERIFICAR a notacao anterior. eq 25 ate aqui}.
{In this point, it is important to highlight that, while the proposed approach parallels classical thermodynamics, it provides a rigorous bridge between quantum and classical descriptions. By employing the Ehrenfest theorem in its thermal average form \eqref{eq:ehrenfest}, we establish a systematic derivation of quantum Maxwell relations \eqref{eq:mx} that naturally allows the incorporation of quantum correlations.}

It is a consensus in the literature that the spin-spin correlation function plays a pivotal role in determining the quantum nature of spin systems \cite{yurishchev2011quantum,aldoshin2014quantum,cruz2016carboxylate,cruz2022quantum,richter2020decay,tiwari2023quantum,turkeshi2023entanglement,fumani2021quantum,cruz2020quantifying}, since the measurement of the statistical correlation between the spins of different particles of the system is directly related to the presence of genuine quantum correlations in the system \cite{yurishchev2011quantum,aldoshin2014quantum,cruz2016carboxylate,cruz2022quantum}. 

For the sake of simplicity, let us consider a dinuclear spin-1/2 system, such as a metal complex in a d$^9$ electronic configuration, providing a nearly perfect representation of an isolated two-qubit system~\cite{cruz2016carboxylate,cruz2017influence,cruz2022quantum,cruz2022,yurishchev2011quantum,aldoshin2014quantum,souza,gaita2019molecular,moreno2018molecular,cruz2020quantifying}. In this case, as shown in references \cite{cruz2016carboxylate,cruz2022quantum}, for any spin-1/2  dinuclear metal complex, %antiparallel aligned ($J\!>\!0$), i.e., with an entangled ground state \cite{cruz2022quantum,cruz2020quantifying}, 
the Schatten 1-norm quantum discord $\Dcal (J,T)$ can be written directly proportional to the  spin-spin correlation function\cite{Ciccarello:14,cruz2016carboxylate,cruz2022quantum}:%can be obtained by the measurement of the system's magnetic susceptibility as \cite{yurishchev2011quantum,aldoshin2014quantum,cruz2016carboxylate,cruz2020quantifying}:
\begin{align}
     \Dcal (J,T) = %=\frac{4}{3+e^{J/ k_B T} }-1 = 
    \frac{1}{2} \left\vert \Ccal (J,T)\right\vert~.%\frac{2 k_B T \chi\left(J, T\right)}{N\left(g \mu_B\right)^2}-1~.
\label{eq:dimero}
\end{align}
where $\Ccal(J,T)= \Ccal_{1,2}^{(x)} =  \Ccal_{1,2}^{(y)}= \Ccal_{1,2}^{(z)}$. For systems with antiparallel alignment ($J > 0$), where the ground state is entangled, the correlation function $\Ccal(J,T)$ ranges between $-1$ and 0. Conversely, for systems with parallel alignment ($J < 0$), where the ground state is separable, $\Ccal(J,T)$ varies from 0 to 1/3. %\textcolor{red}{$\Ccal(J,T)$ nao foi definido, apenas $C_{ij}$ pairwise...} 
Thus, from equation \ref{eq:ehrenfest} we obtain:
\begin{align}
\left\vert \left\langle\frac{d \Hcal}{d J}\right\rangle \right\vert &= 6\Dcal (J,T)~,\label{eq:ehrenfest_correlation}
\end{align}

{
It is worth highlighting that a two-spin-1/2 system with isotropic Heisenberg interaction was chosen as a representative example to illustrate a fundamental link between quantum correlations and the proposed caloric approach. The motivation behind this choice lies in the well-established role of spin-spin correlation functions as measurable indicators of quantum correlations, including quantum discord \cite{cruz2016carboxylate,cruz2017influence,cruz2022quantum,cruz2022,yurishchev2011quantum,aldoshin2014quantum}. It is important to emphasize that our choice of system does not imply that the observed quantum caloric effects are restricted to this particular case. In this regard, the Heisenberg interaction serves as a bench test to demonstrate the potential applicability of our framework. Similar analyses can be extended to other quantum systems in which non-classical correlations. 
}

Therefore the variation of the isothermal entropy, equation \ref{DeltaS_iso}, can be written in terms of the quantum discord of this system as: 
\begin{equation}
\left\vert  \Delta S_{iso}(\Delta J, T) \right\vert= 6\int_{J_A}^{J_B}\left[\frac{\partial \Dcal (J,T)}{\partial T}\right] d J,
\label{DeltaS_iso2}
\end{equation}
In particular, the quantum discord is a monotone decreasing function with temperature for the isotropic Heisenberg interaction \cite{cruz2016carboxylate}. For any system with an entangled ground state ($J>0$), $\Delta S_{iso}(\Delta J, T)<0$, which characterizes the standard caloric effect, it can be associated with a heat rejection of the material since the quantum correlations increase when the coupling is increased. Conversely, for any system with a separable ground state ($J<0$), $\Delta S_{iso}(\Delta J, T)>0$, characterizing the inverse caloric effect, related to a heat absorption of the material when the coupling is reduced, as quantum discord is converted into thermal energy. Thus, the heat exchanges in the material will be dictated by the variation of quantum correlations.

Furthermore, the Schatten 1-norm quantum discord in equation \ref{eq:dimero} can be obtained by the magnetic susceptibility $\chi (J,T)$  \cite{cruz2016carboxylate,cruz2022quantum} as:
{\begin{eqnarray}
			\Dcal=\frac{1}{2}\left| \frac{2 k_B T}{N(g\mu_B)^2}\chi (J,T)-1\right|~,\label{discord2}
			\end{eqnarray}}
where $N$ is the Avogadro's number,  $g$ is the isotropic Land\'{e} factor, and $\mu_B$ represents the Bohr magneton. {Therefore, by selecting a Hamiltonian model in which these quantum correlations are explicitly tied to thermodynamic properties, we provide a concrete and experimentally accessible example of how quantum correlations can influence caloric potentials and heat exchange processes.}

Different from the standard magnetocaloric effect, the measurement of magnetic susceptibility plays a crucial role in analyzing the quantum caloric effect from the perspective of the magnetic coupling constant. Relating to this, one can observe caloric effects throughout low magnetic field measurements. In addition, this measurement is directly linked to the genuine quantum correlations \cite{yurishchev2011quantum,aldoshin2014quantum,cruz2016carboxylate,cruz2020quantifying}, opening a new perspective for studying the caloric properties of advanced materials. 

\begin{figure}[htbp]
    \centering
        \begin{subfigure}{0.45\textwidth}
        \centering
        \includegraphics[width=\textwidth]{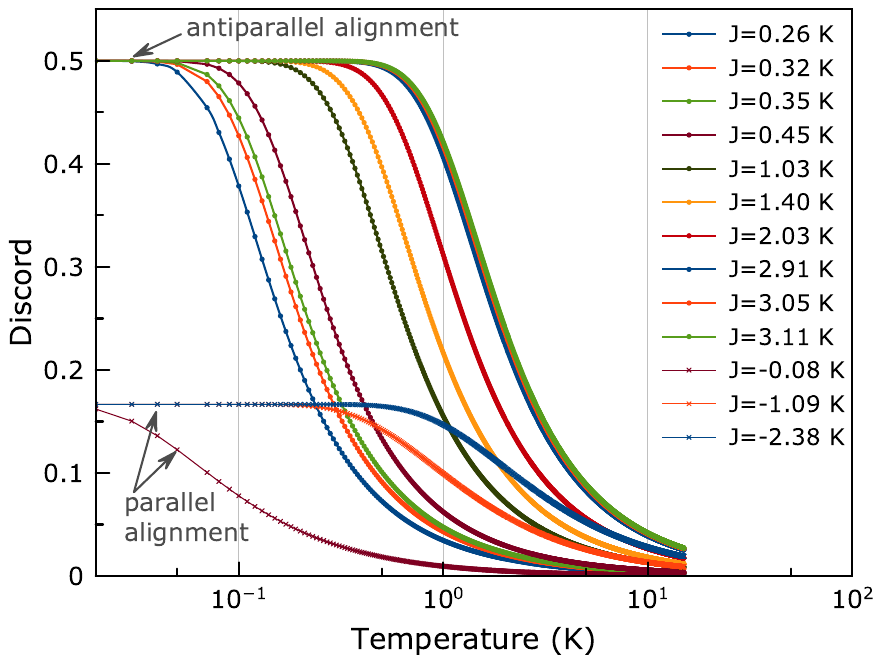}
    \end{subfigure}
    \hfill
    \begin{subfigure}{0.45\textwidth}
        \centering
        \includegraphics[width=\textwidth]{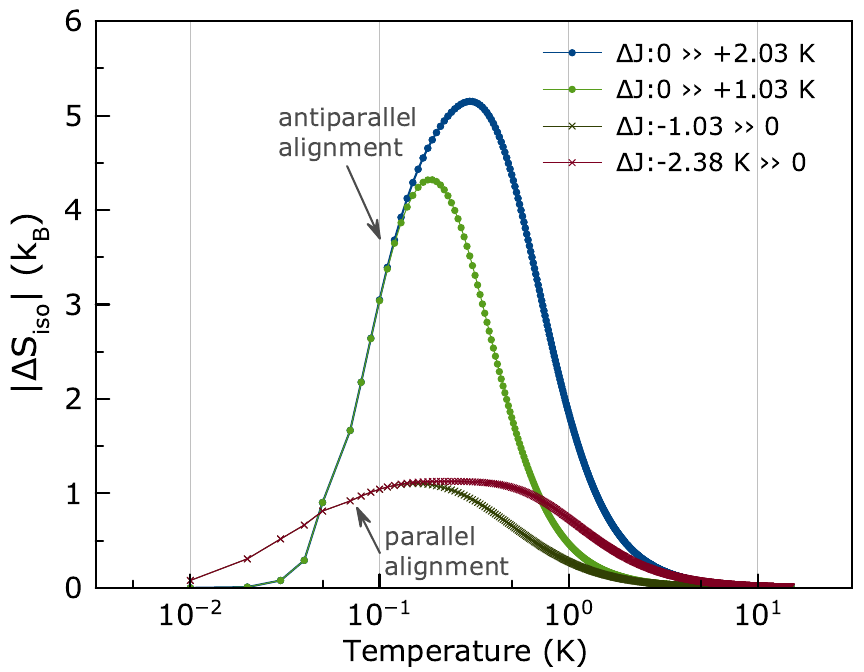}
    \end{subfigure}
\caption{We obtained the exchange parameters from first-principles calculations as a function of pressure  (up to 4.9 GPa) \cite{cruz2017influence} for the prototype material KNaCuSi$4$O${10}$. With these parameters in hands, the corresponding magnetic susceptibility, discord (top panel) and isothermal entropy change (bottom panel) were calculated.}
    \label{fig:dsetc}
\end{figure}

In this regard, we can use magnetic susceptibility curves as a function of the temperature for different values of magnetic coupling to determine the variation of the quantum discord of the system with respect to temperature, allowing us to calculate the isothermal entropy change. 
From an experimental point of view, these curves can be obtained from high-pressure magnetic susceptibility measurements. However, due to the limited availability of a considerable number of susceptibility measurements with different coupling values needed to perform the integration presented in equation \ref{DeltaS_iso2}, we face certain constraints. To circumvent these experimental limitations, we obtained the exchange parameter from first-principles calculations on the prototype material KNaCuSi$_4$O$_{10}$, as a function of pressure (up to 4.9 GPa) \cite{cruz2017influence,cruz2020quantifying}. With these parameters in hands, the corresponding magnetic susceptibility and discord were calculated (see Figure \ref{fig:dsetc}-top). In this material, the degree of quantum correlations in the spin cluster system is managed by changes in magnetic coupling induced by variations in hydrostatic pressure \cite{cruz2017influence,cruz2020quantifying}. Figure \ref{fig:dsetc}-bottom illustrates the isothermal entropy change, equation \ref{DeltaS_iso2}, for this prototype material KNaCuSi$4$O${10}$ \cite{cruz2017influence}. As observed in classical caloric effects, there is a peak in entropy variation as a function of temperature. This peak arises from changes in the system's quantum correlations, which are related to alterations in the population of the material's energy levels due to variations in coupling ($\Delta J$). This indicates the presence of significant caloric effects in this region. Moreover, the peak variation in entropy is greater when the system's ground state is entangled, as such states exhibit more quantum correlations compared to systems with separable ground states. 

{
In this context, it is worth noting that while the role of quantum correlations has been widely explored in quantum thermodynamics, it remains largely unexplored in the context of caloric effects despite a century of studies involving advanced materials \cite{Franco2018Magnetocaloric,reis2020magnetocaloric,DEOLIVEIRA201089}. While caloric effects in quantum materials have been studied extensively \cite{reis2020caloric}, the role of quantum correlations in these processes has received less attention. The classic caloric effect in quantum materials has mainly relied on classical statistical mechanics frameworks \cite{DEOLIVEIRA201089}, where caloric potentials cannot be attributed to quantum discord even when the system Hamiltonian is directly tied to quantum correlations.  In contrast, the presented results show the potential to overcome the classical approach explicitly showing that the isothermal entropy change $\Delta S^{\text{iso}}$ can be directly related to quantum discord, highlighting an intrinsic consequence of non-classical correlations to the energy exchange of advanced materials.

Moreover, the specific example chosen in this work, based on the Heisenberg spin interaction, was selected precisely because it offers a clear demonstration of this principle. By establishing a theoretical link between caloric response functions and quantum correlations, we illustrate the potential of the proposed framework to evaluate the influence of the quantum features of the system on the energy exchange in an isothermal process. It is important to highlight that the formalism developed in this work is general and, although illustrated with a specific example, it can be applied to other quantum systems with a well-defined Hamiltonian model. 
Therefore, this perspective introduces a conceptual contribution to the centenary study of the caloric effects, in which quantum materials are typically treated as classical systems with discrete energy levels. Thus, quantum correlations can be seen as an additional thermodynamic resource for the caloric potentials, and insight remained overlooked until now. In summary, our work contributes to bridging this gap, providing a systematic framework to incorporate genuine quantum features into caloric models.
} It paves the way for further exploration of caloric effects in diverse physical systems and offers potential advancements in quantum refrigeration technologies and quantum thermodynamic applications.

\section{Conclusions} \label{sec5}

In this study, we introduced an unexplored approach by deriving the quantum Maxwell relationship using the thermal average form of the Ehrenfest theorem. This innovative method allowed us to develop comprehensive frameworks for quantum caloric potentials, particularly the isothermal entropy change and adiabatic temperature change. 

Our approach not only extends classical thermodynamic concepts into the quantum realm but also highlights that {the caloric potentia} of quantum systems can be intrinsically linked to genuine quantum correlations, such as quantum discord, {an unexplored aspect of caloric effects that has remained absent on its literature for a century. While the connection between isothermal entropy change and quantum correlations is demonstrated using a system governed by Heisenberg interactions, the underlying thermodynamic relations developed are general. The main insight is that one can evaluate the caloric potential of any quantum system given a well-defined Hamiltonian model. In this regard, our approach remains flexible and can be adapted to various Hamiltonian models, making it a valuable tool for exploring caloric effects in diverse quantum systems.} Conversely, by recovering classical cases within our quantum framework, we demonstrate the robustness and wide applicability of our results, bridging a critical gap between classical and quantum thermodynamics. 

{Therefore, the presented approach contributes to the understanding of the caloric behavior of quantum systems and offers a perspective on how quantum features may influence energy conversion processes, encouraging further studies on the role of quantum entanglement, coherence and other non-classical properties in caloric behavior in different quantum systems.}

	\begin{acknowledgments}
In memory of J. S. de Almeida, whose work was crucial to our research. His influence and generosity are fondly remembered. C. Cruz expresses gratitude to N. G. de Almeida and C. J. Villas-Boas for their valuable comments. C. Cruz thanks the Fundação de Amparo à Pesquisa do Estado da Bahia - FAPESB for its financial support (grant numbers APP0041/2023 and PPP0006/2024). MSR thanks FAPERJ and CNPq for financial support.  MSR belongs to the INCT of Refrigeração e Termofísica,
funded by CNPq by grant number 404023/2019-3. This study was financed in
part by the Coordena\c{c}\~{a}o de Aperfei\c{c}oamento de Pessoal de Nivel
Superior—Brasil (CAPES)—finance code 001. JA declares that this work was developed within the scope of the project CICECO-Aveiro Institute of Materials, UIDB/50011/2020, UIDP/50011/2020 \& LA/P/0006/2020, financed by national funds through the FCT/MCTES (PIDDAC). 

	\end{acknowledgments}
{
\appendix

\section{Thermal analogs of the Ehrenfest Theorem}\label{ehrenfest}

This section provides a detailed calculation of the term:
\begin{equation}\label{eq:esperado}
    \left\langle \frac{d \Hcal (\lambda)}{d \lambda}\right\rangle = \sum_n p_n(\lambda, T) \frac{dE_n(\lambda)}{d\lambda}~.
\end{equation}

Let us start with a general case where both, eigenvalues and eigenvectors of the Hamiltonian depend on a parameter $\lambda$
\begin{equation}
H(\lambda) = \sum_n E_n(\lambda) |n(\lambda)\rangle \langle n(\lambda)|.
\end{equation}
Thus, the derivative of the Hamiltonian with respect to \(\lambda\) is  given by:
\begin{align}\label{eq:derivada}
\frac{d H(\lambda)}{d \lambda} =& \sum_n \Big( \frac{d E_n(\lambda)}{d \lambda} |n(\lambda)\rangle \langle n(\lambda)| + E_n(\lambda) \frac{d |n(\lambda)\rangle}{d \lambda} \langle n(\lambda)| + \nonumber\\ & + E_n(\lambda) |n(\lambda)\rangle \frac{d \langle n(\lambda)|}{d \lambda} \Big).
\end{align}

In thermal equilibrium, the thermal average of an operator ${d H(\lambda)}/{d \lambda}$ is given by:
\begin{equation}
\left\langle \frac{d H(\lambda)}{d \lambda} \right\rangle = \operatorname{Tr}\left[\rho(\lambda,T)\frac{d H(\lambda)}{d \lambda}\right],
\end{equation}
where  $\rho(\lambda,T) = \sum_n p_n(\lambda, T) |n (\lambda)\rangle \langle n(\lambda)|$, $\beta = {1}/{k_B T}$, $k_B$ is the Boltzmann constant, and $T$ is the temperature. Thus,
\begin{equation}
    \left\langle \frac{d H(\lambda)}{d \lambda} \right\rangle = \sum_m \langle m (\lambda) | \left[ \sum_{n} P_{n} | n(\lambda) \rangle \langle n(\lambda)| \frac{d H(\lambda)}{d \lambda}  \right] | m (\lambda)\rangle~.
\end{equation}

Substituting the expression for $\frac{d H(\lambda)}{d \lambda}$ in equation \ref{eq:derivada} and using the orthonormality relation between the eigenstates of the energy basis $\langle m(\lambda) | n(\lambda\rangle = \delta_{mn}$, we have:
\begin{widetext}
\begin{align}
\left\langle \frac{d H(\lambda)}{d \lambda} \right\rangle = 
\sum_n P_n 
\sum_k \Bigg( 
\frac{d E_k(\lambda)}{d \lambda} \langle n(\lambda) | k(\lambda) \rangle \langle k(\lambda) | n(\lambda) \rangle 
+ E_k(\lambda) \langle n(\lambda) | \frac{d |k(\lambda)\rangle}{d \lambda} \langle k(\lambda) | n(\lambda) \rangle 
+ E_k(\lambda) \langle n(\lambda) | k(\lambda) \rangle \frac{d \langle k(\lambda)|}{d \lambda} | n(\lambda) \rangle 
\Bigg).
\end{align}
\end{widetext}

Applying the orthonormality relation $\langle n(\lambda) | k(\lambda)\rangle = \delta_{nk}$, the term ${d E_k(\lambda)}/{d \lambda}$ contributes directly as:
\begin{equation}
\frac{d E_k(\lambda)}{d \lambda} \langle n(\lambda) | k(\lambda) \rangle \langle k(\lambda) | n(\lambda) \rangle = \frac{d E_n(\lambda)}{d \lambda}.
\end{equation}

The cross terms involving derivatives of the eigenstates $|n(\lambda)\rangle$ cancel out due to the orthonormality of the eigenstates. By definition, the derivative of the orthonormality condition:
\begin{equation}
\langle n(\lambda) | n(\lambda) \rangle = 1,
\end{equation}
this implies that:
\begin{equation}
\langle n(\lambda) | \frac{d |n(\lambda)\rangle}{d \lambda} + \frac{d \langle n(\lambda)|}{d \lambda} | n(\lambda) \rangle = 0.
\end{equation}
Thus, each term which involves the derivatives of the eigenstates in the sum cancels individually. Therefore, we have:
\begin{equation}
\sum_n \sum_k \langle n(\lambda) | \frac{d |k(\lambda)\rangle}{d \lambda} \langle k(\lambda) | n(\lambda) \rangle + \langle n(\lambda) | k(\lambda) \rangle \frac{d \langle k(\lambda)|}{d \lambda} | n(\lambda) \rangle = 0.
\end{equation}

This ensures that the total contribution from the derivatives of the eigenstates vanishes and only the diagonal contributions remain significant. Thus, the thermal average of the Hamiltonian derivative is given by:
\begin{equation}
\left\langle \frac{d H(\lambda)}{d \lambda} \right\rangle = \sum_n p_n \frac{d E_n(\lambda)}{d \lambda},
\end{equation}
where $p_n = {e^{-\beta E_n(\lambda)}}/{\sum_k e^{-\beta E_k(\lambda)}}$. This expression depends only on the derivatives ${d E_n(\lambda)}/{d \lambda}$, and the populations $p_n$. 

Therefore, the generalized forge written in the main text can be rewriten as 
\begin{align}\label{eq:ehrenfest0}
Y(\lambda, T) &= -\frac{\dbar \Wcal}{d \lambda} = -\sum_n p_n(\lambda, T) \frac{dE_n(\lambda)}{d\lambda}~,\nonumber\\
&=- \left\langle \frac{d \Hcal (\lambda)}{d \lambda}\right\rangle,
\end{align}
which can be interpreted as a thermal analogs for the Ehrenfest theorem.
}

\end{document}